# The complex dynamics of bicycle pelotons

**Hugh Trenchard**
406 1172 Yates Street
Victoria, British Columbia V8V 3M8
htrenchard@shaw.ca

**Abstract**

A peloton may be defined as two or more cyclists riding in sufficiently close proximity to be located either in one of two basic positions: 1) behind cyclists in zones of reduced air pressure, referred to as 'drafting', or 2) in zones of highest air pressure, described here alternately as 'riding at the front', 'in the wind', or in 'non-drafting positions'. Cyclists in drafting zones expend less energy than in front positions. Two broad models of peloton dynamics are explored. The first is an energetic model that describes peloton dynamics that oscillate through observable phase states as they emerge from collision avoidance and riders' coupled energy outputs. These phases exhibit behavioural characteristics such as convection patterns and synchronization, among others. Under the second, economic model, we discuss some basic parameters of the peloton as a system of economic exchange, and identify the resources within a peloton for which riders compete and cooperate. These include the energy savings of drafting, a near-front positional resource, and an information resource.

## Introduction

The Tour de France is perhaps the best known mass-start bicycle race. It is a race among ~200 riders who compete for top individual and team placing on roads and over distances up to 220km and 23 days (www.letour.fr, 2012).

Mass-start races are also held on oval tracks called velodromes. Standard track lengths are 250m or 333m. Track races consist of multiple laps, and may be between 3km and 40km in distance, composed of perhaps 15 to 40 riders. Mountain bike and cyclo-cross events are also mass-start races, but these are not discussed here, since they usually involve narrow courses on which riders are forced to ride single-file and do not generate the kinds of dynamics explored here.

One characterization of a peloton is that it is the group of riders in a given competition, each of whom employ tactics and strategy for the top positions at the finish line. By this characterization, we may view a peloton as a competitive system composed of riders with multiple objectives that serve to advance riders' overall competitive goals. These objectives may include saving energy by drafting or by cooperating with team-mates or competitors; advancing relative positions nearer to the front; or hiding and seeking information about competitors' positions and their relative energy levels.

In this view, collective peloton behaviours result from the deliberate and calculated competitive objectives of the riders. These behaviours are therefore top-down driven, in the sense that collective behaviours result from deliberately imposed actions by the riders themselves within the peloton, or at the behest of team managers or other external human influences.

An analysis of peloton dynamics on the basis of top-down influences may well enable a sound understanding of cycling as a sport. However, a fuller appreciation of the richness of peloton dynamics is achieved by a conception of a peloton as a complex dynamical system. Under a complex systems approach, certain collective dynamics are self-organized and emerge from physical principles that drive cyclists' local, or nearest neighbour, interactions. These collective behaviours and patterns emerge independently of the riders' deliberate competitive actions, and cannot be predicted by the behaviours of individual cyclists riding in isolation from the group. Further, under a complex systems approach, certain dynamics are also mixed self-organized and top-down in nature.

In this paper, two broad models of peloton dynamics are explored. The first is an energetic model that describes peloton dynamics that oscillate through observable phase states as they emerge from collision avoidance and riders' coupled energy outputs. These phases exhibit behavioural characteristics such as convection patterns and synchronization, among others.

Under the second, economic model, we discuss some basic parameters of the peloton as a system of economic exchange, and identify the resources within a peloton for which riders compete and cooperate. These include the energy savings of drafting, a near-front positional resource, and an information resource.

In the energetic model we are concerned with fundamentally self-organized patterns, while an economic model entails an analysis of a peloton as a mixed system composed of elements that are both top-down and self-organized. Here we hope to set the foundation upon which to build our understanding of the rich collective patterns of behaviour that emerge from these processes.

### Definition

A peloton may be defined as two or more cyclists riding in sufficiently close proximity to be located either in one of two basic positions: 1) behind cyclists in zones of reduced air pressure, referred to as 'drafting', or 2) in zones of highest air pressure, described here alternately as 'riding at the front', 'in the wind', or in 'non-drafting positions'. Cyclists in drafting zones expend less energy than in front positions. These zones are located either directly behind or beside at angles to other cyclists, depending on wind direction. For large pelotons (approx. >6), proportionately more cyclists will be in drafting positions than in front positions.

Energy expenditure when drafting a single rider is reduced by approximately 18% at 32km/hr (20mph), 27% at 40km/hr (25mph), and by as much as 39% at 40km/hr in a group of eight riders (McCole et al, 1990). At the elite level, speeds of 40 to 50km/hr on flat topography are common, and pelotons of 100 or more cyclists are common. Because there is an approximate energy savings of 1% per mph when riding behind one ride (Hagberg and McCole 1990) (Figure 1), for convenience speeds here are sometimes shown in miles per hour, as well as in metric values. For discussion on the factors to be considered in calculating more precisely the drafting benefit for variable wheel spacing between riders, and the effects of peloton size on speed, see Olds (1998).

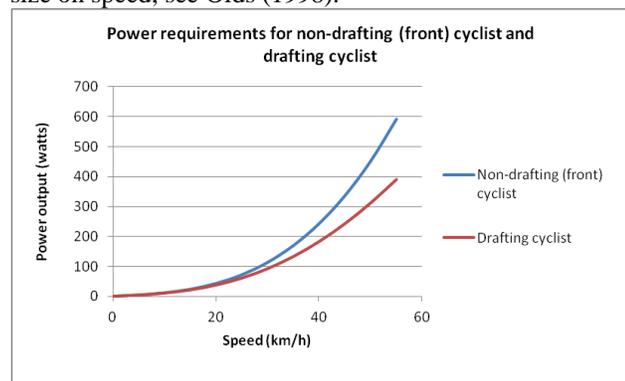

**Figure 1.** Power requirements for cyclist in non-drafting position and cyclist in drafting position. Curve for non-drafting cyclist based on 75kg (bicycle and rider); rolling friction coefficient 0.004 dimensionless; 0.00 gradient; air-density 1.226kg/m$^3$; drag co-efficient of 0.5; frontal surface area of 0.05m$^2$ (parameters from www.analyticcycling.com). Curve for drafting cyclist based on approximate 1% savings per mile/hr (Hagberg and McCole 1990; Burke, 1996; Figure adapted from Trenchard, 2011).

## An Energetic Model

Coupling occurs between cyclists when one or more seek the energy-saving benefits of drafting, while simultaneously adjusting positions to avoid collisions. A cyclist's power requirement to overcome wind resistance is proportional to the cube of his or her velocity (Burke, ed. 1996). In order to overcome wind resistance, approximately one percent of total energy expenditure required to overcome wind-resistance is reduced per one mile an hour by drafting behind a single cyclist, while greater reductions occur by riding in the middle of a larger pack (Hagberg and McCole 1990), although below approximately 10mph, drafting benefit is negligible (Swain 1998; Figure 1).

Cyclists' power output is not determined only by speed; it may vary according to position (drafting or non-drafting), riders' speed being equal. Also, speed falls in proportion to the slope of the road (Swain 1998), while power output may remain constant. Conversely, speed may be high on a descent, but power output low.

### Drafting as the basis of peloton cohesion

By taking advantage of the energy savings benefits of drafting, cyclists' energy expenditures/power outputs are thus coupled, and by alternating peloton positions to optimize energy expenditures, cyclists in groups can sustain speeds at lower power outputs than individuals riding alone (Olds, 1998). Thus by drafting, cyclists in a group effectively equalize the differences in their power output capacities. This equalization effect is the basis for tactics and strategy in bicycle racing as cyclists seek to overcome this effect.

It is well understood among cyclists that the energy savings benefit of drafting facilitates peloton cohesion. Nonetheless, from an empirical standpoint, it is beneficial to use data to show peloton formation as a function of drafting. For this we compare five time-trial finishing time distributions with five finish time distributions of mass-start races.

In time-trials, competitors commence at timed intervals and are not permitted to draft. Time-trial data from five time-trials (Figure 2) reveals that, for these races, distribution of finishing times is Gaussian. Time-trialing ability is a strong indicator of a cyclist's capacity to keep pace within a peloton, though it is not the only one.

Acceleration capacity and bicycle-handling skill are others.

In mass-start races groups of riders finish so near to each other that it is impractical for time-keepers to allocate time differences between them, and they recorded as finishing at the same time. Further, under international cycling sport regulations, when riders finish in a given bunch all of them must be credited with the same finishing time (UCI Regulation 2.3.040, 2010), and as such the difference in times between these riders is zero.

From the perspective of collective dynamics, the recording of these identical times is more than merely practical or regulatory, but reflects the aggregate and collective nature of the peloton.

Figure 2 shows this clustering effect is typically observed in mass-start cycling races; it is clearly an effect of the cohesive nature of drafting, as indicated by Olds (1998) and well understood among cyclists, and as expounded here.

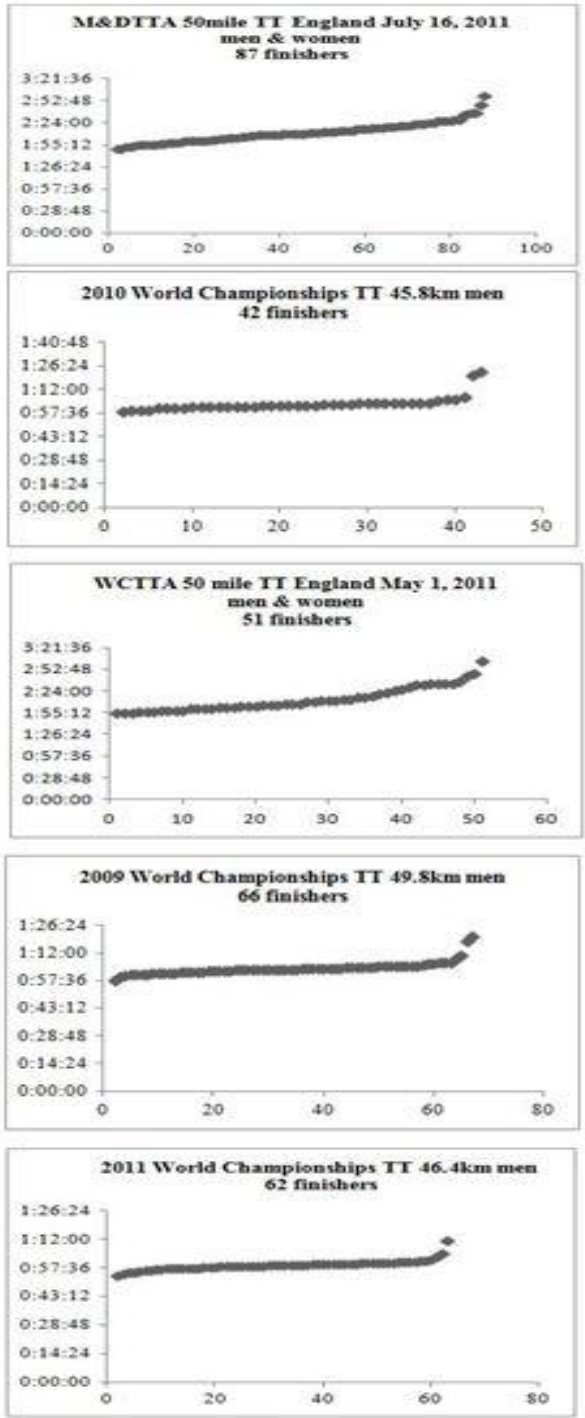
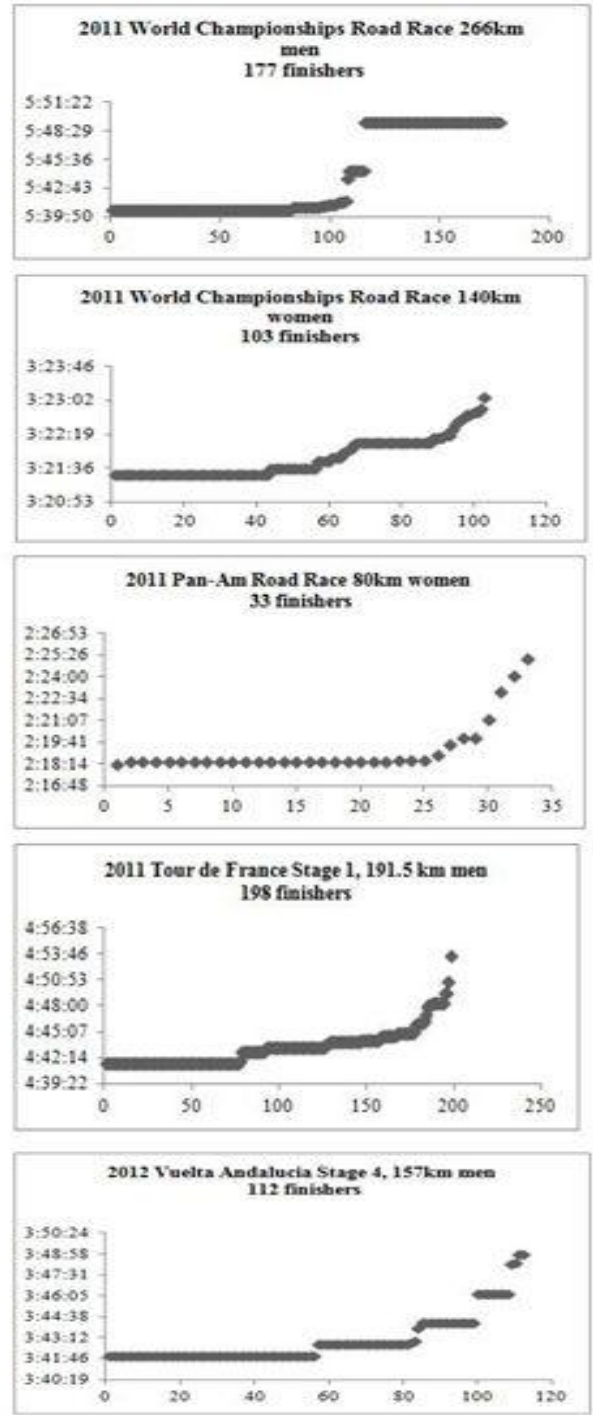

**Figure 2.** Finish time distributions for five individual time-trials and five road races. Finish times are plotted on the y-axis; individual cyclists are plotted on x-axis. Time-trials, for which drafting is not permitted under race regulations, show a characteristic Gaussian distribution. Road races, for which drafting is permitted, show characteristic cluster distributions, where riders who finish as a group receive the same finishing time, and groups are well separated. It is well understood among cyclists that drafting represents the "glue" that sustains high peloton density, and cyclists employ a variety of tactics to overcome this factor.

**Coupling Description** Coupling capacity between two cyclists is well described by the ratio of the difference between power outputs of two cyclists at any given time, and the drafting advantage for speeds at those times. A simplified occurrence where maximum sustainable outputs between riders are offset by the drafting advantage, referred to here as the Peloton Divergence Ratio (PDR) (adapted from Trenchard, H. and Mayer-Kress, 2005) is given by:

**PDR = ((Wa – Wb) / Wa) / (D/100)**

Where **Wa** is the maximum sustainable power output (watts) of cyclist A at any given moment; **Wb** is the maximum sustainable power of cyclist B at any given moment (assuming Wa>Wb); D/100 is the percent energy savings (correlating to reduced power output) due to drafting at the velocity travelled. This is called a divergence ratio because it reveals the critical power output threshold at which coupled riders de-couple (Table 1). More succinctly, PCR shows the proportion of strength a weaker rider, B, must be to a stronger one, A, before B cannot keep up to A when drafting.

Here "maximum sustainable output" refers to outputs sustainable for specific times to exhaustion as a fraction of $V0_{2max}$ (Olds 1998), or maximum volume of oxygen consumed in litres/min. $V0_{2max}$ is an expression of physiological capacity for work and frequently referred to among cyclists. Thus, the lower the fractional $V0_{2max}$, the longer the time to exhaustion.

As a guide, consistent with the definitions above of event duration length, a sprint of maximal anaerobic output (90% anaerobic) is sustained for less than ten seconds, while a sub-maximal effort (50/50 anaerobic/aerobic requirement) is sustainable for approximately two minutes; an aerobic output (99% aerobic) may be sustained for hours (McCardle et al., 2006).

The discussion in this paper will generally involve maximal sustainable efforts >10s and <2min, which are known durations a front rider will take his turn in the wind. However, the maximum sustainable effort can be scaled to apply to all threshold maximums which may occur momentarily or over longer durations. We refer to this threshold also as an "athletic fitness threshold" or "competitive fitness threshold".

| Riding speed (km/h) | 20 | 32 | 40 | 50 |
|---|---|---|---|---|
| Estimated reduction (%) | 18 | | | 32 |
| Measured reduction (%) | | 18 | 27 | |
| Approximate incline gradient for maximal power output at speeds noted | 8 | 4 | 2 | 0 |
| Example rider A: max power output in each situation (watts) | 450 | 450 | 450 | 450 |
| Example rider B: max power output in each situation (watts) | 420 | 420 | 420 | 420 |
| PDR = ((Wa-Wb) / Wa) / (D/100) | .07/.18 (0.39) | .07/.18 (0.39) | .07/.27 (0.26) | .07/.32 (0.22) |
| Example rider C: max power output in each situation (watts) | 390 | 390 | 390 | 390 |
| PDR =((Wa-Wc) / Wa) / (D/100) | .13/.18 (0.72) | .13/.18 (0.72) | .13/.27 (0.48) | .13/.32 (0.41) |
| Example rider D: max power output in each situation (watts) | 360 | 360 | 360 | 360 |
| PDR = ((Wa-Wd) / Wa) / (D/100) | *.20/.18 (1.11)* | *.20/.18 (1.11)* | .20/.27 (0.74) | .20/.32 (0.63) |
| Example rider E: max power output in each situation (watts) | 330 | 330 | 330 | 330 |
| PDR = ((Wa-We) / Wa) / (D/100) | *.27/.18 (1.50)* | *27/.18 (1.50)* | *.27/.27 (1.00)* | .27/.32 (0.84) |

**Table 1.** Example coupling ratios, applying PDR. Estimated and measured reduction in energy expenditure at various speeds (Kyle, C. 1979; McCole, S.D. et al, 1990; Burke, E., 1996), approximate corresponding slope gradients for maximal power outputs at those speeds, and four examples of coupling ratios. Values are rounded. Above PDR threshold 1 (bold italics) peloton divergences occur. Below this threshold, riders are coupled. Power output largely dependent on body weight and other factors; approximations here based on same values as in Figure 1. (Table adapted from Trenchard, H. and Mayer-Kress, G., 2005).

A modified, more powerful version of PDR is referred to here as the Peloton Coupling Ratio (PCR). The nominative difference is arbitrary, but to distinguish them we call the first a divergence ratio because it models necessary conditions for a weaker rider to keep pace with a stronger front rider, as well as the conditions which lead to divergence; PCR, on the other hand, indicates how a weaker rider can sustain convergence by changing from a front position to a drafting position, or repeated alternations of this process. Further, under PCR, we know the proportion that the following rider's output is to her maximum and her required output to maintain convergence, whereas PDR tells us only whether the following rider's actual maximum sustainable output is sufficient to maintain pace with the front rider.

PCR is given by:

$$PCR = [(WaMa) - ((WaMa)*(D/100))] / Wa$$

Where:
- **Wa** is *both*[1] the maximum sustainable power output (watts) of the following cyclist, and the threshold output pace set by the front rider; any speed set by the front rider over this threshold, and the rider's de-couple; in other words the front rider may be capable of yet higher outputs than **Wa;**
- **Ma** is the proportion of the rider's current speed to her maximum sustainable maximum speed when not drafting (**Ma = Current speed / max sustainable speed**). **Ma** is based on speed to be consistent with drafting advantage;
- **D** is the percent energy savings from drafting, approximated by 1% per mph (Hagberg and McCole 1990; Burke 1996).

[1] See *Appendix 1* for an alternative representation

In order to derive D, it is not enough to know only power output, but we must also know the speeds at which the riders travel. Hence, multiplying **Wa** by **Ma** we obtain the equivalent power output proportion of current speed to maximum sustainable speed.

Thus **(WaMa) – [(WaMa) * (D/100)]** provides the *required output* of the following rider while drafting to maintain the speed set by the front rider. As in our final equation, PCR, when following riders' required output exceeds her maximum sustainable output as set by the front rider, the riders de-couple (Table 2). This of course occurs when the front rider accelerates to a speed and output over the threshold **Wa**. In cycling parlance the following rider is thus "dropped", or "off the back".

Since the two riders are coupled as to speed (but not power output) which is set by the front cyclist (Fig 5b), the front cyclist's speed is therefore incorporated into the equation. Further, as indicated, the power output of the front cyclist is reflected by **Wa$^1$**, which is both the maximum sustainable output of the following rider and the threshold output of the front rider before de-coupling (i.e. if the front rider increases output over that threshold, the riders de-couple). Hence the equation shows whether or not the following rider is capable of keeping pace with the speed set by the front rider while exploiting the power reduction benefit of drafting.

Note the output of the front, non-drafting rider, will always be greater than the current output of the following rider at the given speed because D is not factored into the front rider's output.

Applying both PDR and PCR we can see that a significantly stronger rider whose maximum sustainable speed is, for example, 25mph relative to a following rider whose maximum sustainable is 20mph, may never be able to drop the drafting rider even if they are changing positions. This is a simplified situation consisting of unchanging wind-direction, flat course topography and the ability of the following rider to maintain optimal drafting position. As noted, in practice course conditions are constantly changing, and the stronger rider can drop the following rider by forcing the following rider into non-optimal drafting positions through short rapid accelerations ("attacking") either from the front or from behind, or switching positions rapidly across the road exposing the following rider temporarily to the wind, or by taking advantage of reduced drafting opportunities on hills when speeds are slower, but power output may remain high. These are all tactics employed by riders to overcome the benefits of D.

When there are many riders involved, the same tactics are employed, but they are distributed across many riders and shared among team-mates, who may alternate attacks.

Under a complex systems framework, we may think of these tactics as creating instabilities in an equilibrium coupled state.

The PCR is illustrated in Table 2 and Figure 3. Figure 3 incorporates data from Table 2, and represents a fundamental diagram of two riders accelerating through a range of 16mph to 29mph, and where the weaker rider, B, begins in front and rides to her maximum sustainable speed, and then changes positions with the stronger rider, A.

Figure 3 shows that when riders change positions where stronger rider A is in front, the PCR curve breaks to a higher range. Also, the fitness differences among riders diminish the more elite the race. Hence, when the fitness range among riders is comparatively narrow, as it is in elite races, riders' tactics involve continuous attempts to create coupling instabilities.

The process of changing positions may repeat continuously, and as long as PCR ≤1, the riders remain coupled. Selected maximum speeds for cyclist A is 29mph, and for B, 20mph. In real world circumstances, the course gradients, wind speeds and directions are constantly changing; and because riders' sustainable maximum speeds and D change correspondingly, actual data for this model would be difficult to acquire.

Also note that although Figure 2 represents a range of speeds of the weaker rider up to her maximum sustainable speed when not drafting (switching with the stronger rider at that point), in practice riders will usually switch positions before they reach their maximum.

Although the equations are similar to the extent that they both predict divergence, for PCR, unlike PDR, the front rider is not required by the equation to be stronger than the following rider; it expresses coupling degrees between cyclists of different strengths, who may be either in front or behind, and who may be riding at varying proportions of their maximum sustainable outputs.

Both PDR and PCR indicate that divergence between cyclists *necessarily* results when the maximum output of the following cyclist is less than the output of the rider ahead minus the drafting benefit. Under the conditions of the equation, this will never occur when a stronger rider is drafting behind a weaker rider, who enjoys the double advantage of drafting and being stronger. It should be noted, however, that in reality there are times when a stronger, following rider, relaxes, loses concentration, or encounters mechanical difficulties sufficient for the weaker rider to slip away. Examples of this are when the following rider is drinking or eating, or crashes or punctures, and the front rider rides away.

| A | B | C | D | E | F | G | H | I |
|---|---|---|---|---|---|---|---|---|
| \multicolumn{9}{c}{**Cyclist A (stronger) in drafting position**} ||||||||||
| current speed (mph) coupled cyclists A & B, as set by front rider | max speed A | D/100 | Wa | Ma = current speed/ max speed | WaMa | (WaMa)* (D/100) | WaMa – ((WaMa)* (D /100)) | PCR = [(WaMa) – ((WaMa)* (D/100))] / Wa |
| 16 | 29 | 0.16 | 658* | 0.55 | 362 | 58 | 304 | 0.46 |
| 17 | 29 | 0.17 | 658 | 0.59 | 388 | 66 | 322 | 0.49 |
| 18 | 29 | 0.18 | 658 | 0.62 | 408 | 73 | 335 | 0.51 |
| 19 | 29 | 0.19 | 658 | 0.66 | 434 | 82 | 352 | 0.53 |
| 20 | 29 | 0.20 | 658 | 0.69 | 454 | 91 | 363 | 0.55 |
| \multicolumn{9}{c}{*Cyclists change position*} ||||||||||
| \multicolumn{9}{c}{**Cyclist B (weaker) in drafting position**} ||||||||||
| current speed (mph) coupled cyclists A & B, as set by front rider | max speed B | D/100 | Wa | Ma = current speed/ max speed | WaMa | WaMa* (D/100) | WaMa – ((WaMa)* (D/100)) | PCR = [(WaMa) – ((WaMa)* (D/100))] / Wa |
| 21 | 20 | 0.21 | 345** | 1.05 | 362 | 76 | 286 | 0.83 |
| 22 | 20 | 0.22 | 345 | 1.10 | 380 | 84 | 296 | 0.86 |
| 23 | 20 | 0.23 | 345 | 1.15 | 397 | 91 | 306 | 0.88 |
| 24 | 20 | 0.24 | 345 | 1.20 | 414 | 99 | 315 | 0.91 |
| 25 | 20 | 0.25 | 345 | 1.25 | 431 | 108 | 323 | 0.94 |
| 26 | 20 | 0.26 | 345 | 1.30 | 449 | 117 | 332 | 0.96 |
| 27 | 20 | 0.27 | 345 | 1.35 | 466 | 126 | 340 | 0.99 |
| 28 | 20 | 0.28 | 345 | 1.40 | 483 | 135 | 348 | *1.01* |
| 29 | 20 | 0.29 | 345 | 1.45 | 500 | 145 | 355 | *1.03* |

**Table 2**. Hypothetical data for two coupled cyclists increasing speed from 16mph to 29mph.

**Column A.** Rider B is in front position for speeds 16mph to 20mph when B reaches her maximum power output, while A is drafting. Riders change position so B is drafting for speeds 21mph to 29mph. Due to drafting, B can continue increasing speed up until just over 27mph, when PCR>1.
**Column B.** Arbitrarily set maximum speed for cyclist A of 29mph without drafting, and max sustainable power output for of 658W; for cyclist B, arbitrary maximum speed is 20mph without drafting, and output of 345W.
**Column C.** D is approximate percent energy savings due to drafting at speed travelled.
**Column D.** Wa is the following rider's maximum output as it corresponds to his maximum speed, and the threshold output of the front rider before causing de-coupling.
**Column E.** Ma is a ratio of current speed to maximum sustainable speed without drafting. Note for cyclist B, this ratio exceeds 1 because the required power when not drafting is higher than his sustainable maximum; required power is reduced by drafting however, the magnitudes of which are shown in column D, and he is thereby capable of sustaining higher speeds as long as PCR <1, or in this example, 27mph, as shown in column I. This requirement for cyclist B to draft in order to maintain speeds above 20mph is indicated by the central row "cyclists change position".
**Column F.** WaMa is required power output at current speeds for both riders without drafting, using ratio of current speed to max speed.
**Column G.** shows power output saving due to D.
**Column H.** shows current reduced power output due to D.

*power 658W for 29mph calculated using www.analyticcycling.com: 12.964m/s; frontal area 0.50m$^2$; coefficient wind drag 0.50 dimensionless; air density 1.226kg/m$^3$; weight cyclist and bicycle 75kg; rolling resistance 0.004 dimensionless; grade 0.030 decimal; crank length 170mm.

** power of 345W at 20mph calculated using www.analyticcycling.com: 9.13m/s; frontal area 0.50m$^2$; coefficient wind drag 0.50 dimensionless; air density 1.226kg/m$^3$; weight cyclist and bicycle 75kg; rolling resistance 0.004 dimensionless; grade 0.030 decimal; crank length 170mm.

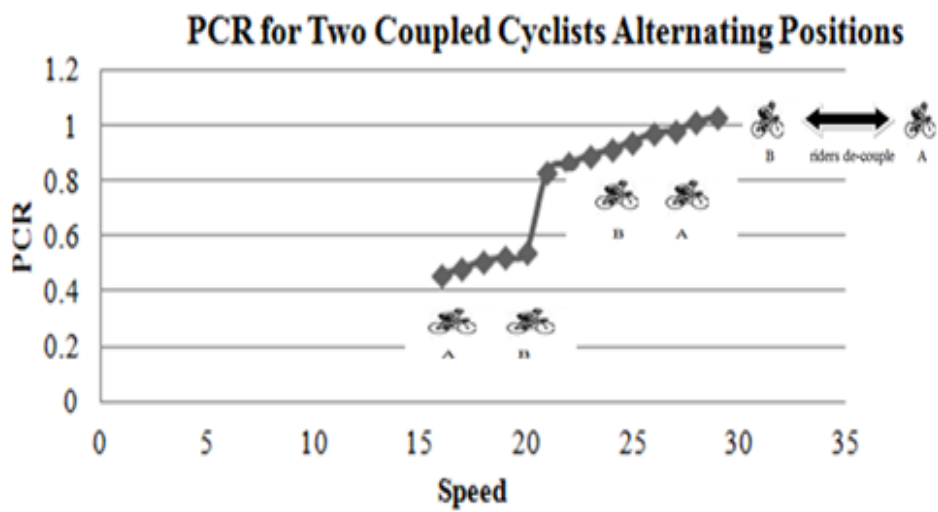

**Figure 3**. Hypothetical data in Table 2 yields the fundamental diagram for two coupled cyclists where a weaker rider begins in front and proceeds to maximum speed and output, and who then changes position with a stronger rider, allowing the weaker rider to draft and to continue increasing speed beyond her maximum speed capable without drafting. Thus over speeds 16mph to 20mph, weaker rider B is in front while stronger rider A is drafting. When B reaches maximum output at 20mph, riders change positions so A is in front and B is drafting. PCR is in lower range when stronger rider drafts, and curve jumps to higher range when they switch positions. When D no longer offsets the difference between front (stronger) rider's output and the following (weaker) rider's maximum, PCR > 1 and riders de-couple.

## Global coupling and the emergence of complex dynamics

The foregoing describes coupling effects between two riders. As an aggregate of more than two riders, all cyclists within a peloton are at PCR < 1 relative to each other, a further defining element of a peloton. Greater drafting occurs among central riders (McCole, et al., 1990), and varying degrees of drafting occur in the lateral direction, depending on wind direction and speed. Riders adjust positions for optimal drafting position and to avoid collision; every movement affects the movements of others.

As in the case of an individual rider falling out of drafting range of another rider or peloton, when a group of coupled riders de-couples, forming a sub-peloton, and decelerates relative to the main peloton containing the largest number of riders, separated pelotons are at PCR > 1, measured by the output of the front rider of the sub-peloton to the last rider in the main peloton, and assuming increasing divergence at the same or greater rate after separation.

PCR thus allows complex peloton dynamics and phase states and their transitions to be identified according to PCR ranges.

## Phase States

**Phase I Relaxed.** When riders assemble at the start line of a mass-start race without forward motion and with zero power output, they form a static precursor state to Phase I that is not significant to this analysis. However, immediately upon commencement of motion from stationary positions, Phase I dynamics begin. As cyclists begin forward motion and accelerate, their power outputs are well below sustainable maximums, resulting in PCR

<1. Because riders proceed at below maximum outputs in this phase, riders have an abundance of energetic resources which are used inefficiently. As such distances between riders are not optimized for maximum drafting benefit and a comparatively high proportion of riders make only minor positional adjustments which do not drive an increase in peloton speed, as they do in Phase II. In addition to their occurrence at the outset of a race, Phase I dynamics may also be observed as collective relaxations following high output accelerations which may be of sufficient aggregate intensity to reach a Phase IV (PCR>1) state. In comparatively short races, a Phase I state may exist only for the brief duration required for cyclists to accelerate from zero to their sustainable maximums, and may never be observed afterward.

**Phase II Convection Rolls.** In this phase, peloton rotations occur – a situation in which riders advance up peloton peripheries as riders in central positions fall effectively toward the back (Fig. 4), forming a convection dynamic whereby groups of "warming" cyclists advance up the peripheries, and "cooling" cyclists effectively shift backward along central positions in the peloton.

This phase is described as a convection roll, similar to convection currents in fluid (Rayleigh, 1916) as greatest energy output ("warming") occurs on peloton extremities, while decreasing output occurs effectively backward through central peloton positions ("cooling"; Fig 4). It is suggested here that this roll dynamic is a function of two factors: the expenditure of stored energy as rested cyclists pass fatiguing riders who have just expended relatively greater quantities in non-drafting positions (a self-organized dynamic); and a front-position imperative, whereby riders endeavor to maintain close proximity to the front, which is a mixed self-organized/top-down dynamic, as also discussed subsequently under the economic model.

The front-position imperative arises because of the competitive advantage offered by these positions. As riders advance up the peripheries, their positional change causes riders who are passed effectively to be shifted to rear positions in the peloton, despite their frequent attempts to hold their front positions. Those who attempt to maintain central-internal positions near the front by passing within the peloton are largely prevented by riders ahead who block the way. When the effectively backward movement occurs, those riders shifted back are naturally motivated to advance again (Fig. 4).

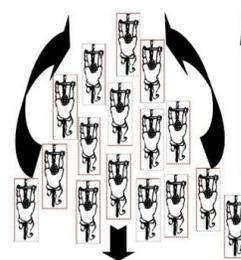

**Figure 4.** Peloton convection roll illustration. Long straight arrow indicates direction of peloton. Curved arrows indicate general rotation: riders pass up peloton peripheries. Notched arrow indicates effective direction of riders in the middle of the peloton as they are passed by riders moving up peripheries.

**Phase III Synchronization.** In a peloton, cyclists synchronize speed and phase-lock power output when their speed is increased to a critical theshold at which riders self-organize into a paceline, whereby cyclists ride one immediately behind another (Trenchard and Mayer-Kress, 2005).

Phase-locking occurs when the motion frequencies of two or more oscillators are identical but are shifted in the origin of the output oscillation, or are out-of-phase (Strogatz and Stewart, 1993). For a cyclist, output oscillations are his variations in power output. Here "motion frequency" equates with rider velocity. It is important to distinguish between self-organized pacelines and those that form as a result of agreement during training rides or those that are deliberately formed for team time-trials, for example, which we are not concerned with here.

When cyclists' power outputs are phase-locked, the greatest difference in power output is between the front rider and any of the following riders, such that all following riders' power outputs are phase-locked to the first rider in a paceline (Fig 5b).

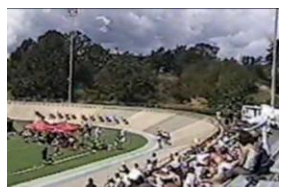 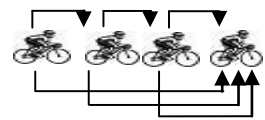

**Figure 5 a.** Speed synchronization and phase locking of riders on a velodrome (photo by the author); **b.** Illustration of three following riders (FR), whose power outputs are phase-locked to a pacesetter (farthest right of diagram). Each FR is phase-locked primarily to the pacesetter who sets speed (lower arrows) at greatest power output and secondarily to other FR from whom they receive direct drafting advantage (upper arrows)

Pacelines may self-organize among subsets of riders and and do not necessarily involve the entire peloton. It is suggested here that pacelines self-organize when pacesetters (PS) set a threshold speed which following riders (FR) are incapable of matching and sustaining without drafting (PCR approaches 1). In this formation, pacesetters cannot sustain their effort for as much time as drafting riders, and must reduce their speed allowing a following rider to take the front position; this process continues.

In a paceline, PS modulates the effective wind speed, and PS's power output is substantially higher than FR in drafting positions. When PS decelerates, he may do so simply by "pulling off to the side", as noted, allowing following riders to maintain phase-locking and consistency of speed, at which time a new rider becomes PS, and FR phase-locked synchrony is maintained among remaining paceline riders. If PS and FR decelerate approximately simultaneously, synchrony breaks down as FR adjust positions and speeds to avoid collision. If PS accelerates beyond the threshold speed so that PCR > 1, phase-locking is also destroyed as FR cannot match the acceleration. This effect is easily observed on inclines where there is reduced drafting advantage, when cyclists proceed nearer to individual maximum sustainable outputs without drafting (Table 1).

**Phase IV Disintegrated.** As noted, this phase is visually similar to Phase I dynamics in terms of its low density, but the two phases are generated from opposite extremes of power outputs. Phase IV disintegration occurs when PCR>1 in conditions of low drafting benefit, such as on hills, or when riders efforts are near sustainable maximums during an intermediate attack, or during a finishing sprint. Disintegration can occur in isolated regions of the peloton, and are often of comparatively short duration except in cases of long ascents of sufficiently steep slope. In the cases of short term and local disintegrations, Phase IV states are mixed with Phases II and III.

As observed previously, unless Phase IV states occur at the finish of a race or the race is of comparatively short duration, they are usually followed by a relaxation (Phase I) state, in which cyclists' outputs decrease suddenly and by a comparatively large magnitude.

**Peloton Hysteresis.** Peloton hysteresis may be modeled in part upon analyses of vehicle traffic hysteresis (Trenchard, 2010). A succinct description of hysteresis in traffic flows (Kuhne and Michalopolous, 1997) is as follows:

**The dynamics of traffic flow result in the hysteresis phenomena. This consists of a generally retarded behaviour of vehicle platoons after emerging from a disturbance compared to the behavior of the same vehicles approaching the disturbance**.

In vehicle traffic analysis, density is defined as vehicles per hour per traffic lane involving a continuous passage of vehicles past designated points of measurement (Taylor, 2005). In a peloton, density is similarly described except for the presence of a finite number of riders. Peloton density may be thus described as a flow parameter of the time required for a set number of riders to pass a specific point, such as a start finish line on a track (Trenchard, 2010).

Peloton hysteresis appears to occur on courses in which a peloton approaches an incline (hill) in which drafting advantage decreases, and while cyclists decelerate into corners and accelerate out of them (the "accordion effect") (Fig. 6). A third kind of "competitive hysteresis" is indicated by velodrome mass-start data (Trenchard, 2010), in which cyclists accelerate as density falls, after which cyclists gradually decelerate as density continues to fall (i.e. spread continues to increase) to a threshold point when density begins to increase again.

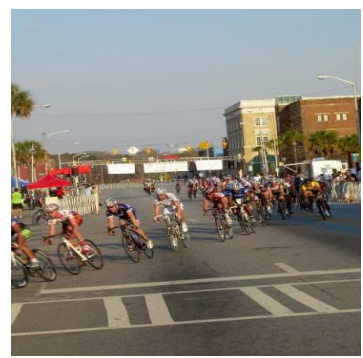

**Figure 6.** A criterium and the accordion effect. Riders adjust to take single tangent through corner, requiring some to decelerate as they approach the corner, increasing peloton density leading into the corner. As riders exit the corner, they accelerate and decrease density (photo by the author).

## Foundations of an Economic Model

In this part, the intention is only to introduce the application of economic concepts to peloton dynamics. To that end, we identify the scarce resources for which cyclists compete, and thereby lay a foundation upon which a broader economic model may be constructed.

Further, the economic problem is broadly outlined, but it is not within the scope of this analysis to weigh costs and benefits in detail, or to model formally the resource exchange transactions or cooperative relationships that comprise the structure of an economic model.

Beginning with the basic premise that economics involves competition for scarce resources (McConnell, et al., 2005), cyclists must weigh the costs and benefits of seeking certain peloton resources (subsequently defined), and choose their best options accordingly, thus engaging in a network of economic transactions. To that end, we may model the costs and benefits of assumed rational actors (cyclists) in a competitive system of resource exchange as they advance their competitive goals.

So far we have looked at peloton dynamics that are predominantly self-organized, emerging from basic physical principles of collision avoidance and energy savings by position optimization. An economic analysis, however, encompasses both self-organized and top-down dynamics. Cyclists' decisions that are based on cost/benefit analyses and imposed as acts of their own volition are top-down in nature, compared with responsive actions or ones dictated by physiological limitations, which are bottom-up. Hence an economic model of the peloton must take into account the combined nature of the self-organized processes already discussed, and the volitional choices cyclists make in advancing their competitive objectives.

## Peloton resources

There are at least three primary resources within a peloton for which cyclists compete. The most significant of these is the energy savings offered by the drafting, as discussed. The second resource is less tangible but nevertheless important in the context of a mass-start bicycle race: close proximity (CP) to the front of the peloton, alternatively described earlier as a front-position imperative. The third peloton resource is information: cyclists alter their competitive responses according to information they acquire about the positions or apparent fatigue of other riders.

**Drafting Resource.** A reduction in energy expended is a physically tangible resource as it is experienced directly through riders' physiological feedback (i.e. they can "feel" their effort is easier when drafting), and the benefits of drafting (D) significantly outweigh the energetic costs of acquiring it, given a mass-start race in which riders have these resources immediately and abundantly available right from the start, and the cost is lies in *maintaining* D, rather than having to seek it from the start. The benefit of maintaining D even outweighs the high costs associated with frequent short term maximal efforts (sprints or extended periods above anaerobic threshold) required to stay within drafting range of others when attacks occur. Collectively, such efforts are rewarded by higher average speed, particularly in the event of breakaways, and the formation of sub-groups, and continued access to D at higher speeds.

**Front Position Resource.** The second resource is less tangible than the first, but nevertheless important in the context of a mass-start bicycle race: close proximity (CP) to the front of the peloton. CP does not include non-drafting positions at the front of the peloton, but includes positions *near* the front that allow riders simultaneously to draft as well as remain in tactically advantageous positions to respond to other cyclists' attacks (quick and sudden accelerations), or the final sprint for the finish.

For CP, cyclists compete for a limited number of optimal near-front positions, balancing the trade-off between the comparatively high energetic costs in achieving these positions against the energy savings benefits of drafting. Optimal positions are located at or near the front of the peloton, where breakaway attempts may be launched, or from which the winner ultimately emerges – the farther back a cyclist is, the less likely it is that he or she can win the race. Experienced cyclists explain that cyclists should try to stay in the top ten to 15 places - positions for which pelotons composed of well over 100 riders may compete.

Because peloton peripheries represent the fastest route for achieving CP, cyclists seek peripheral positions in order to advance. These peripheries, however, entail comparatively greater wind exposure while also requiring cyclists to accelerate in order to pass others and hence requiring increased energy expenditure in their efforts to advance.

By contrast, maximum energy savings are achieved by drafting within the peloton. However, as a natural consequence of riders advancing up peripheries, riders located within the peloton are inevitably shifted nearer to the back of the peloton (a component of Phase III convection dynamics, as discussed earlier). As a result, cyclists must continually weigh the high energy costs of advancing up peloton extremities to gain positional advantage, against the energy savings of internal maximum drafting positions.

The benefits of advancing must also be weighed against the costs of passing riders when separations occur among them (PCR>1) at positions farther back in the peloton, or against the costs of advancing position when cyclists' own drafting advantage is lost due to a significant change in course gradient or effective changes in wind direction, or when course constraints make forward advancement impossible or too costly to attempt.

It is noteworthy that these costs are frequently borne in part by team-mates, known as domestiques, whose roles are specifically to pace team leaders to the front, to close separations or maintain a minimum distance to breakaway riders (riding tempo, as previously mentioned), to provide drafting opportunities in cross-winds, or to take maximum effort/duration pulls when in breakaways with a team leader or when approaching the race finish.

CP is an example of a positional resource as discussed by Morrell and Romey (2008), who refer to research in which animal collectives must balance the trade-off between the risk of predation and the opportunities to consume food resources at peripheral positions in the collective. Predators are shown to select prey at collective peripheries, including: lapwings, (Salek and Smilauer, 2002); spiders (Rayor and Uetz, 1990); mussels (Okamura 1986); shoals of fish (Bauman et al. 1997). By contrast, Morrell and Romey (2008) also refer to research that shows the benefits of peripheral positions in the consumption of food: burrowing spiders (Lubin et al. 2001); colonial spiders (Rayor and Uetz 1990, 1993); groups of whirligig beetles (Romey, 1995). In this view, there are evolutionary implications for peloton dynamics, further details of which are not explored in this discussion.

**Information Resource.** Peloton information comprises three main categories: displayed information, hidden information (Trenchard, 2011) and hidden information obtained by information systems (Gueguen, 2007).

Displayed information is generally available to all riders, although it is often obscured. Displayed information includes: rider positions and collective configurations, their movement patterns; time-gaps between groups, rider speeds, the course profile and its constraints and obstacles. Displayed information also includes visible properties of each cyclist, such as their body mass and general physical appearance, facial expression, color of uniform, type of bicycle, gear selections, riding style, or quantity of liquid in bottles.

Hidden information is available only to each individual rider unless voluntarily shared, or involuntarily expressed through body signals to reveal hunger and thirst, relative strength, degree of suffering, quantity of food in pockets, among other things. Generally riders do not voluntarily share accurate information about fatigue with opponents. However, in certain situations, such as when riders alternate positions in the wind, riders may indicate temporary fatigue by gesturing with their elbow or a flick of one hand, or by simply decelerating and allowing themselves to be passed by fresher riders. Elbow/hand gestures may not accurately indicate rider fatigue, however, as riders may deliberately exhibit inaccurate signals by bluffing.

Hidden information obtained by information systems includes radio links between cyclists and team managers (headphones), TV screens in team manager cars, GPS localization for instant measures of gaps between rider, power output monitors and heart rate monitors (Gueguen, 2007). Although largely available to the riders directly, this information is often monitored more closely by team managers than by riders, who are preoccupied with the racing environment. Managers may analyze the information and relay directions back to the riders by radio.

The value of information varies, measured in terms of the energy riders are willing to expend to obtain it. For example, the energy value of an opponent's position is greater than the energy value of the amount of water in his own water bottle, and riders will expend comparatively large quantities of energy to learn of others' positions. This situation may be reversed, however, if it is a very hot day and the rider has nothing left to drink. In this case, the rider may seek the best position in which to acquire a full water bottle and be less concerned about the positions of opponents, at least until his bottle is refilled. In all cases, information acquisition is weighed as a cost against its tactical value in advancing riders competitive objectives.

## Conclusion

As an aggregate of interacting heterogeneous units (cyclists), a peloton is a complex dynamical system exhibiting self-organized and mixed self-organized/top-down emergent behaviours. We have examined a few such behaviours, and have identified coupling by drafting as the fundamental unit that underlies peloton dynamics. We have developed an energetic model applying coupling principles and have identified self-organized phase states through which a peloton oscillates.

We have presented the foundation of an economical model composed of three main peloton resources. In this view, a peloton is a network of continuous economic exchange as cyclists constantly evaluate the various costs and benefits involved in acquiring these resources, and take action or respond accordingly.

Ultimately, peloton dynamics are driven by human volitional motivations as well as by natural physical forces. From this combination emerges a rich diversity of complex behaviour that is ripe with insight into our understanding of a variety of human social and economic dynamics, as well as the behaviours of other biological systems.

## Acknowledgements

The author acknowledges posthumously the input of Gottfried Mayer-Kress on the PDR equation and synchronization aspects of this paper; the input and editing of Demian Seale, and Kristin Scott.

# Appendix 1

$$PCR = 1/ [Wa / (WaMa - (WaMa*D/100))]$$

This is an alternative equation that does not require **Wa** to constitute two equivalent characterizations. Here the value 1 represents the output of the front rider as equivalent to the maximum sustainable power output of the drafting rider, which is **Wa**; i.e. the output of the front rider cannot exceed 1 in the coupled state.

The benefit of D allows the following rider to proceed at lower than required outputs (as required without benefit of D), up to her maximum sustainable power output. Thus where PCR>1, the following rider's maximum sustainable power output is insufficient to sustain the speed set by the front rider, since D is not sufficient to offset the required output of the following rider at the given speed, and the riders will de-couple.